\documentclass{elsart}
\usepackage{graphics}
\usepackage{graphicx}
\usepackage{psfig}
\usepackage{color}

\def\qed{\hfill  \framebox(5,5){}}

\def\pp{{\mathcal  P}}
\def\cc{{\mathcal  C}}

\def\sig{{\rm sign}}

\expandafter\chardef\csname pre amssym.def
at\endcsname=\the\catcode`\@ \catcode`\@=11

\def\undefine#1{\let#1\undefined}
\def\newsymbol#1#2#3#4#5{\let\next@\relax
 \ifnum#2=\@ne\let\next@\msafam@\else
 \ifnum#2=\tw@\let\next@\msbfam@\fi\fi
 \mathchardef#1="#3\next@#4#5}
\def\mathhexbox@#1#2#3{\relax
 \ifmmode\mathpalette{}{\m@th\mathchar"#1#2#3}%
 \else\leavevmode\hbox{$\m@th\mathchar"#1#2#3$}\fi}
\def\hexnumber@#1{\ifcase#1 0\or 1\or 2\or 3\or 4\or 5\or 6\or 7\or 8\or
 9\or A\or B\or C\or D\or E\or F\fi}

\font\tenmsa=msam10 \font\sevenmsa=msam7 \font\fivemsa=msam5
\newfam\msafam
\textfont\msafam=\tenmsa \scriptfont\msafam=\sevenmsa
\scriptscriptfont\msafam=\fivemsa
\edef\msafam@{\hexnumber@\msafam} \mathchardef\dabar@"0\msafam@39
\def\dashrightarrow{\mathrel{\dabar@\dabar@\mathchar"0\msafam@4B}}
\def\dashleftarrow{\mathrel{\mathchar"0\msafam@4C\dabar@\dabar@}}

        \font\tenmsb=msbm10
\font\sevenmsb=msbm7 \font\fivemsb=msbm5
\newfam\msbfam
\textfont\msbfam=\tenmsb \scriptfont\msbfam=\sevenmsb
\scriptscriptfont\msbfam=\fivemsb
\edef\msbfam@{\hexnumber@\msbfam}
\def\Bbb#1{\fam\msbfam\relax#1}

\newtheorem{theorem}{{\bf Theorem}}
\newtheorem{remark}{{\bf Remark}}
\newtheorem{definition}[theorem]{{\bf Definition}}
\newtheorem{corollary}[theorem]{{\bf Corollary}}
\newtheorem{proposition}[theorem]{{\bf Proposition}}
\newtheorem{lemma}[theorem]{{\bf Lemma}}
\newtheorem{example}{{\bf Example}}

\begin{document}
\begin{frontmatter}

\title{Local Shape of Generalized Offsets to Algebraic Curves}

\author{Juan Gerardo Alcazar\thanksref{proy}},
\ead{juange.alcazar@uah.es}

\address{Departamento de Matem\'aticas, Universidad de Alcal\'a,
E-28871-Madrid, Spain}

\thanks[proy]{Author supported by the Spanish `` Ministerio de
Educaci\'on y Ciencia" under the Project MTM2005-08690-C02-01.}

\begin{abstract}
In this paper we study the local behavior of an algebraic curve under a geometric construction which is a variation of the usual offsetting construction, namely the {\it generalized} offsetting process (\cite {SS99}). More precisely, here we discuss when and how this geometric construction may cause local changes in the shape of an algebraic curve, and we compare our results with those obtained for the case of classical offsets (\cite{JGS07}). For these purposes, we use well-known notions of Differential Geometry, and also the notion of {\it local shape} introduced in \cite{JGS07}.
\end{abstract}
\end{frontmatter}

\section{Introduction}

The notion of {\sf generalized offset} (see \cite{Juani-thesis}, \cite {SS99} for a more formal definition of this notion and a large study of algebraic and geometric properties) arises in the literature as a generalization of usual offsets. In order to introduce this notion, one may consider the
following construction over a given algebraic curve ${\mathcal
C}$: for every non-isotropic, regular point $P\in {\mathcal C}$, take the
normal line $L_P$ to ${\mathcal C}$ at $P$, rotate it $\theta$
degrees, and consider the points $P_{\pm d,\theta}$ lying on $L_P$
at a distance $d$ of $P$. Then the generalized offset ${\mathcal
G}_{d,\theta}({\mathcal C})$ is the Zariski closure of the set
consisting of all the points $P_{\pm d,\theta}$ computed this way.
In this context, the usual notion of offset (which corresponds to
the case when $\theta$ defines a rotation leaving $L_P$ invariant)
is called the {\sf classical offset} ${\mathcal O}_d({\mathcal
C})$ of the curve (see \cite{Farin}, \cite{HL97}, \cite{PP98b}). For example, in Figure 1 one has, for different
distances, the classical and the generalized offsets for
$\theta=\pi/4$ of an ellipse. Notice that this construction works
over ${\Bbb C}$; nevertheless, in the following we will assume that
${\mathcal C}$ is real, and we will focus on the real part of its generalized
offset, for a real distance and a real angle.

\begin{figure}[ht]
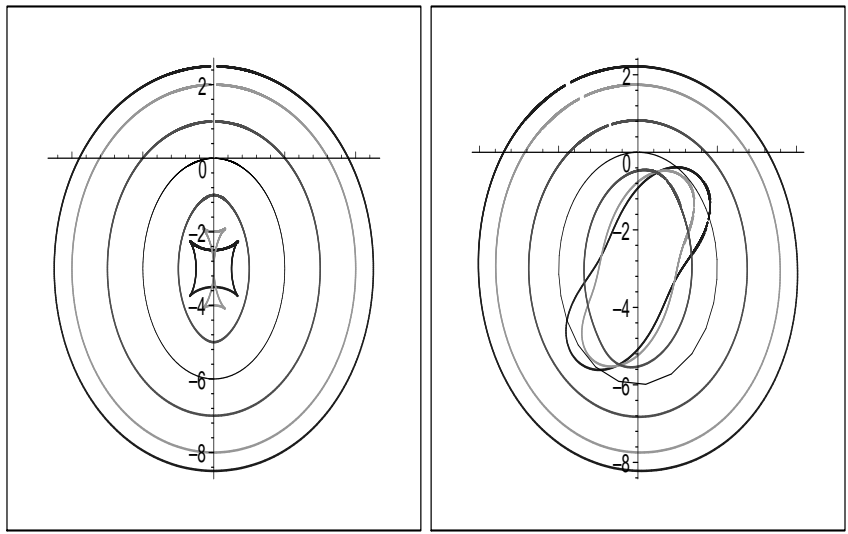

\begin{center}
\centerline{$\begin{array}{lc}
\psfig{figure=Generalized_elipse02.eps,width=5.5cm,height=7cm} &
\psfig{figure=Generalized_elipse03.eps,width=5.5cm,height=7cm}
\end{array}$}
\caption{Classical (left) and Generalized (right) offsets to the
ellipse for different distances}
\end{center}
\end{figure}

Algebraic properties of generalized offsets have been considered in the literature (see  \cite{ASS96}, \cite{ASS97}, \cite {SS99}, \cite{SS00}). In this sense, a nice result is
that properties like the number of components, genus and therefore rationality, are invariant for the angle $\theta$; so, they are shared by all the generalized offsets (including the classical offset) of a given curve. Thus, it is natural to wonder whether the same happens when the shape of generalized offsets is considered.
This paper explores this problem from a local point of view.

Questions on the shape of classical offsets have already been analyzed (see \cite{JGglobal}, \cite{JGS05}, \cite{JGS07}, \cite{Faroukki}, \cite{Far2}). Moreover, in \cite{JGS07} local aspects on the shape of classical offsets of possibly singular algebraic curves are studied. In that paper the notion of {\it local
shape} is introduced in order to
locally describe the shape of a curve. Basically, this notion describes the shape of a real branch of an algebraic curve in the vicinity of a point. So, one may prove (see \cite{JGS05}, \cite{JGS07}) that there are four different behaviors that a real branch can exhibit, which can be found in Figure 2 (see Section \ref{sec-prelim}), corresponding to so-called {\it local shapes (I), (II), (III), (IV)}. Moreover, each of these possibilities has a characterization in terms of {\it places} (see also Section \ref{sec-prelim} in this paper; for more information on the notion of place, we refer the reader to \cite{walker}). Hence, given a geometric transformation like classical or generalized offsetting, in order to analyze how the transformation locally affects the curve one can take a generic place, compute the places it gives rise to in the transformed object, and compare the local shapes of the original and the final places. If all these local shapes coincide, then it means that the transformation has not introduced local changes in the shape of the curve; otherwise, some local change has occurred. Since this strategy can be applied at both regular and singular points, in particular the notion of local shape gives us a way of analyzing the behavior at singularities. The notion of local shape has also been used in \cite{JGglobal} for addressing not only
local, but also global questions on the shape of classical offset curves.


Finally, the structure of the paper is the following. In Section \ref{sec-prelim} we provide the necessary background for developing our results; in particular, the notion of local shape is reviewed here. In Section \ref{sec-behav-reg} we address the behavior of regular points under generalized offsetting processes; the results in this section are proven by using elements of Differential Geometry, without making use of the notion of local shape. In Section 4, we use the notion of local shape for giving a more complete description of the phenomenon, including the behavior at singularities. In Section \ref{sec-comparison}, we summarize the main results in the paper and we provide a comparison between the local properties of the shapes of classical and non-classical generalized offsets.

{\it Acknowledgements.} The author wishes to thank J. Rafael Sendra for suggesting the problem.

\section{Local Shape of an Algebraic Curve} \label{sec-prelim}


In the following we work with an algebraic curve ${\mathcal C}$ different from a line, a real distance $d\neq 0$, and a real angle $\theta$. One may easily see that generalized offsets to lines are also lines; therefore, for lines the analysis is trivial. Since ${\mathcal C}$ is algebraic, around every real non-isolated point $P\in {\mathcal C}$ one can find at least one
local parametrization ${\mathcal P}(h)=(x(h),y(h))$ where $x(h),y(h)$ are real analytic functions and $P={\mathcal P}(0)=(x(0),y(0))$. In the language of {\it places}  (see \cite{walker}) one says that $P$ is the {\it center} of the {\it place} ${\mathcal P}(h)$. The functions $x(h),y(h)$ are called the {\it coordinates} or the {\it components} of the place, and are analytic in a neighborhood $I$ of $0$. Now writing \[x(h)=a_0+a_1h+a_2h^2+\cdots,y(h)=b_0+b_1h+b_2h^2+\cdots,\]we represent by $\mbox{ord}_x$ the {\it order} of $x(h)$, i.e. the least non-zero power of $h$ in the expression of $x(h)$; similarly we introduce $\mbox{ord}_y$. Moreover, we speak of ``real" places to denote places where the coefficients $a_0,a_1,\ldots,b_0,b_1,\ldots$, perhaps after a change of parameter, are real numbers. Then we consider the following definition.

\begin{definition} \label{pairassociated}Let
${\mathcal P}(h)$ be a real place of $\cc$. The {\sf signature} of
${\mathcal P}(h)$ is defined as the pair $(p,q)$ where $p$ is the
first non-zero natural number such that the derivative $\pp^{(p)}(0)\neq \vec{0}$,
and $q>p$ is the first natural number such that
$\pp^{(p)}(0),\pp^{(q)}(0)$ are linearly independent. We denote by
$\sig(\pp(h))$ the signature of $\pp(h)$.
\end{definition}

Since ${\mathcal C}$ by hypothesis is not a line, the numbers $p,q$ in Definition \ref{pairassociated} always exist. Now if $\sig(\pp(h))=(1,q)$ then we say that $\pp(h)$ is {\it regular}, otherwise we say that it is {\it singular}. The center of a singular place is always a singular point of ${\mathcal C}$. Now, in \cite{JGS07} (see Proposition 3 there) it is proven that in a suitable coordinate system, every real non-isolated point $P\in {\mathcal C}$ is the center of a real place $\pp(h)=(x(h),y(h))$ of the type $\pp(h)=(h^p,\beta_qh^q+\cdots)$ where $(p,q)$ is the signature of the place. If a place has this form, we say that it is in {\it standard} form; notice that when the place is in standard form, $\mbox{ord}_x=p<\mbox{ord}_y=q$. Furthermore, in \cite{JGS05}, \cite{JGS07} it is shown that the local behavior of ${\pp(h)}$ around its center can be read from the signature, giving rise to the notion of {\sf local shape}. We recall this notion here.

\begin{definition}\label{def1}
Let ${\mathcal P}(h)$ be a real place of signature $(p,q)$,
centered at $P\in \cc$. Then we say that:
\begin{itemize}
\item[(1)]  ${\mathcal P}(h)$ is a {\sf thorn} (or  it has {\sf local
shape} {\sf (I)}) if  both $p,q$ are even.
\item[(2)] ${\mathcal P}(h)$ is an {\sf elbow} (or  it has {\sf local
shape} {\sf (II)}) if $p$ is odd, and $q$ is even.
\item[(3)] ${\mathcal P}(h)$ is a {\sf beak} (or it has {\sf local
shape} {\sf (III)}) if $p$ is even, and $q$ is odd.
\item[(4)] ${\mathcal P}(h)$ is a {\sf flex} (or  it has {\sf local
shape} {\sf (IV)}) if both $p,q$ are odd.
\end{itemize}
\end{definition}

In Figure 2 one can see the shape corresponding to each local
shape up to rotations. In each case, the center of the place is
the intersection point of the two dotted lines. Furthermore, in
all cases the horizontal dotted line is tangent  to  $\cc$ in the
direction of  $\pp^{(p)}(0)$. We also note  that if ${\mathcal
P}(h)$ is regular, then $p=1$,
 and therefore the only possibilities for the local shape of
${\mathcal P}(h)$ are (II) or (IV). Moreover, if $p$ is even we say that the place is {\it cuspidal}.

\begin{figure}
\begin{center}
  \includegraphics[width=8 cm, height=6 cm]{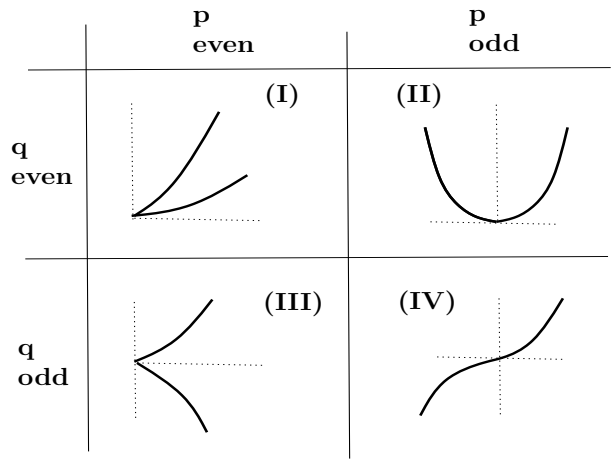}\\
  \caption{Local Shapes}
\end{center}
\end{figure}


\section{Behavior at regular points} \label{sec-behav-reg}

In the rest of the paper, we will represent the generalized offset of ${\mathcal C}$, for a distance $d\neq 0$ and an angle $\theta$, as ${\mathcal G}_{d,\theta}({\mathcal C})$; in particular, if $\theta=0,\pi$ we have the classical offset, ${\mathcal O}_d({\mathcal C})$. Moreover, for local aspects in the topology of classical offsets we refer the reader to \cite{JGS07}, \cite{Faroukki}, \cite{Far2}. So, here we focus on generalized, non-classical, offsets. Now along this section  let ${\mathcal P}(h)=(x(h),y(h))$ be a real regular place of ${\mathcal C}$. Since ${\mathcal P}(h)$ converges in a neighborhood $I$ of $0$, we can regard $(x(h),y(h))$, with $h\in I$, as the parametrization of a regular curve; moreover, we can assume that it has been reparametrized by the arc-length. We will represent by $\bar{r}(h)$ the vector whose components are the coordinates of ${\mathcal P}(h)$. Furthermore, we denote by $\bar{r}_0(h)$ the vector whose components are the coordinates of a place generated by ${\mathcal P}(h)$ in ${\mathcal G}_{d,\theta}({\mathcal C})$. Hence, denoting as $\bar{n}$ the normal vector to ${\mathcal P}(h)$ at its center (i.e. the normal vector to the curve represented by ${\mathcal P}(h)$ at the point $P_0={\mathcal P}(0)$) and denoting the matrix defining a rotation of angle $\theta$ as $A$, it follows that \[\bar{r}_0=\bar{r}+dA\bar{n}\]Now the first result, which shows an important difference between classical and generalized offsets, is the following.

\begin{theorem} \label{th-no-cusps}
The only generalized offset which may transform a regular place into a singular offset place, is the classical offset. Therefore, the generalized, non-classical, offset, never generates a cusp from a regular point of the original curve.
\end{theorem}

{\bf Proof.} Differentiating the equality $\bar{r}_0=\bar{r}+dA\bar{n}$ w.r.t. the arc-length and using Frenet equations, it follows that \[\bar{r}_0'=(I+dkA)\cdot \bar{r}'\]where $k$ is the curvature of ${\mathcal P}(h)$ at its center. Now $\bar{r}_0'=\vec{0}$ iff $\bar{r}'\in \mbox{Ker}(I+dkA)$. However, $\mbox{det}(I+dkA)=(1+dkcos\theta)^2+d^2k^2(sin\theta)^2$. Then $\det(I+dkA)=0$ iff $1+dkcos\theta=0$ and simultaneously $dksin\theta=0$. Since we are assuming that $d\neq 0$ and $k\neq 0$ (i.e. ${\mathcal C}$ is not a line) this holds iff $sin \theta=0$, i.e. when one is working with the classical offset, and $k=-1/d$. In particular, if the offset is non-classical then $\mbox{Ker}(I+dkA)=\{\vec{0}\}$; since we start from a regular place, then $\bar{r}'\neq \vec{0}$ and therefore $\bar{r}_0'\neq \vec{0}$. \qed

\begin{remark} \label{tangents}
When the offset is classical, it is well-known that the tangents to the curve and its offset are parallel at corresponding points. For the generalized offset, the above expression $\bar{r}_0'=(I+dkA)\cdot \bar{r}'$ tells us that this no longer happens; moreover, the tangent line to the generalized offset at a point $Q$ is not even the $\theta$-rotation of the tangent line to ${\mathcal C}$ at the point $P$ generating $Q$.
\end{remark}

In Figure 3 one may see, for $d=1$, the classical offset to the parabola $y=x^2$, and a detail of this offset showing two cusps; in Figure 4 one has the generalized offset of the same curve, also for $d=1$ and a very small angle, $\theta=\pi/50$. The reader may see in Figure 4 that in the generalized offset the cusps have been replaced by rounded arcs.

\begin{figure}[ht]
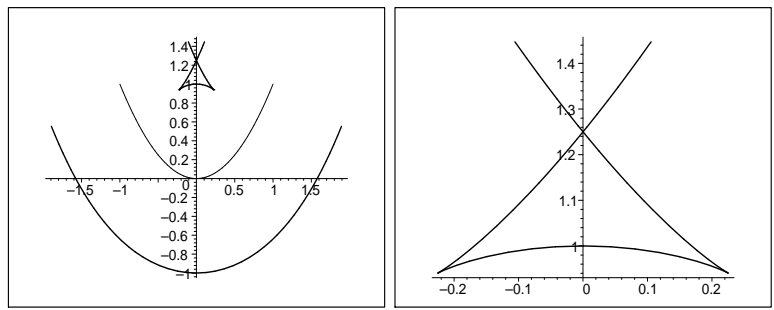

\begin{center}
\centerline{$\begin{array}{ccc}
\psfig{figure=Generalized_parab_tiny04.eps,width=5cm,height=4cm} &
\psfig{figure=Generalized_parab_tiny03.eps,width=5cm,height=4cm} &
\end{array}$}
\end{center}
\caption{Classical Offset to the parabola $y=x^2$, $d=1$ (left); detail (right) }
\end{figure}

\begin{figure}[ht]
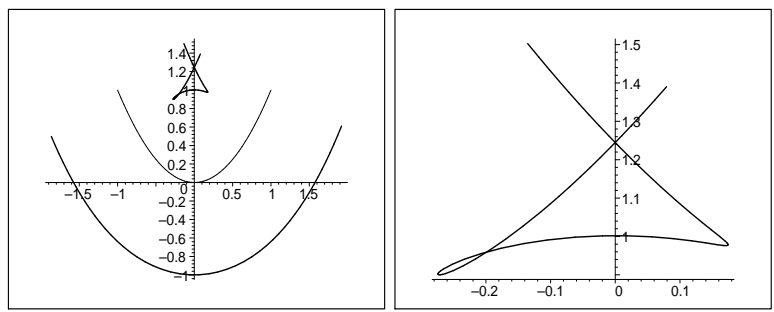

\begin{center}
\centerline{$\begin{array}{ccc}
\psfig{figure=Generalized_parab_tiny02.eps,width=5cm,height=4cm} &
\psfig{figure=Generalized_parab_tiny01.eps,width=5cm,height=4cm} &
\end{array}$}
\end{center}
\caption{Generalized Offset to the parabola $y=x^2$, $\theta=\pi/50$, $d=1$ (left); detail (right)}
\end{figure}

Now let us address the question of checking whether the local shape of regular places is preserved or not by the generalized offsetting process (we say that the local shape of a place is {\it preserved}, if the local shapes of the places that it generates in the generalized offset coincide with the original local shape). Since regular places are either elbows or flex points, the question reduces to analyzing whether generalized offsets preserve flex points coming from regular places. For the classical offsets the answer is ``yes" (see \cite{Faroukki}); however, in the generalized, non-classical case, we will see that the answer is ``no". For this purpose, we recall that the curvature at a regular flex point is $0$. Hence, let $k_0$ denote the curvature of ${\mathcal G}_{d,\theta}({\mathcal C})$ at the center of the place generated by ${\mathcal P}(h)$; from the well-known formula of the curvature, we have that \[k_0=\displaystyle{\frac{(\bar{r}'_0\times \bar{r}_0'')\cdot \bar{z}}{|\bar{r}_0'|^3}}\]where $\bar{z}=(0,0,1)$ is normal to the plane containing $\bar{r}'_0$ and $\bar{r}''_0$. Thus, the following theorem holds.

\begin{theorem} \label{th-reg-flex-points}
The regular points of ${\mathcal C}$ generating flex points of ${\mathcal G}_{d,\theta}({\mathcal C})$, satisfy \[dk'sin\theta+k(k^2d^2+2dkcos\theta+1)=0\]As a consequence, the generalized, non-classical, offset does not necessarily preserve flex points.
\end{theorem}

{\bf Proof.} Let us compute the numerator of the above expression for $k_0$. In order to do this, we have that $\bar{r}_0'=(I+dkA)\bar{r}'=\bar{r}'+dk\cdot A\bar{r}'$. Differentiating again, we get \[\bar{r}_0''=\bar{r}''+dk'A\bar{r}'+dkA\bar{r}''\]Thus, \[\bar{r}_0'\times \bar{r}_0''=dk'\bar{r}'\times A\bar{r}'+\bar{r}'\times \bar{r}''+dk\bar{r}'\times A\bar{r}''+dk A\bar{r}'\times \bar{r}''+d^2k^2 A\bar{r}'\times A\bar{r}''\]Notice that $k'$ (i.e. the derivative of the curvature w.r.t. the arc-length) exists because since ${\mathcal P}(h)$ is regular, then $k$ is an analytic function. Now since $A$ represents a rotation of angle $\theta$ then $|A\bar{r}'|=|\bar{r}'|=1$ (because we are assuming that ${\mathcal P}(h)$ has been re-parametrized w.r.t. the arc-length), and $|A\bar{r}''|=|\bar{r}''|$. Moreover for the same reason the angle between, on one hand, the vectors $A\bar{r}', \bar{r}'$, and on the other hand, the vectors $A\bar{r}'',\bar{r}''$, is $\theta$. Furthermore, if we represent by $\alpha$ the angle between $\bar{r}'$ and $\bar{r}''$, the angle between $A\bar{r}''$ and $\bar{r}'$ is $\alpha + \theta$, and similarly the angle between $A\bar{r}'$ and $\bar{r}''$ is $\alpha-\theta$. Hence,
\[
\begin{array}{ccc}
\bar{r}_0'\times \bar{r}_0'' & = & dk'\sin\theta \cdot \bar{z}+\bar{r}'\times \bar{r}''+dk\cdot |\bar{r}'||\bar{r}''|\sin(\alpha+\theta)\bar{z}+\\
& & + dk\cdot |\bar{r}'||\bar{r}''|\sin(\alpha-\theta)\bar{z}+d^2k^2\bar{r}'\times \bar{r}''
\end{array}
\]Now, expanding $\sin(\alpha+\theta)$ and $\sin(\alpha-\theta)$, taking into account the formula for $k$ in terms of $\bar{r}', \bar{r}''$ and $\bar{z}$, and computing the dot product with $\bar{z}$, one gets that
\[(\bar{r}'_0\times \bar{r}''_0)\cdot \bar{z}=dk'sin\theta+k(k^2d^2+2dkcos\theta+1)\]
Now from Theorem \ref{th-no-cusps} it holds that $|\bar{r}'_0|\neq 0$, and hence $k_0=0$ iff $dk'sin\theta+k(k^2d^2+2dkcos\theta+1)=0$; then, every point of ${\mathcal C}$ giving rise to a flex point of the generalized offset fulfills this equality. Finally, notice that a regular flex point of ${\mathcal C}$ satisfies that $k=0$, but not necessarily that $k'=0$. So, such a point does not necessarily fulfills the condition in the statement of the theorem and therefore flex points are not necessarily preserved. \qed

In fact, in the next section we will see that generalized, non-classical, offsets {\it never} preserve flex points (see Corollary \ref{no-flex-points}). Also, observe that the condition in Theorem \ref{th-reg-flex-points} is not sufficient because the fact that $k_0=0$ does not necessarily imply that the point in ${\mathcal G}_{d,\theta}({\mathcal C})$ is a flex (it depends on the order of the first non-vanishing derivative of $k_0$).

Finally, we address the {\it turning points} (i.e. points of either horizontal or vertical tangent) of the generalized, non-classical offset. In the classical case, it is well-known that the tangents to ${\mathcal C}$ and ${\mathcal O}_d({\mathcal C})$ at corresponding points, are parallel; hence, turning points of the offset are generated by turning points of the original curve, and conversely. However, the following result shows that for generalized, non-classical, offsets this property does not hold in general.

\begin{theorem} \label{th-turning-points}
Let ${\mathcal G}_{d,\theta}({\mathcal C})$ denote a generalized, non-classical offset of ${\mathcal C}$. The following statements are true:
\begin{itemize}
\item [(1)] The points of ${\mathcal G}_{d,\theta}({\mathcal C})$ with vertical tangent, generated by regular points of ${\mathcal C}$, correspond to: (i) points of ${\mathcal C}$ with vertical tangent, where $k=0$; (ii) points of ${\mathcal C}$, with $k\neq 0$, where the slope of the tangent equals $\displaystyle{-\frac{1+dkcos\theta}{dk sin \theta}}$.
    \item [(2)] The points of ${\mathcal G}_{d,\theta}({\mathcal C})$ with horizontal tangent, generated by regular points of ${\mathcal C}$, correspond to: (i) points of ${\mathcal C}$ with horizontal tangent, where $k=0$; (ii) points of ${\mathcal C}$, with $1+dkcos\theta=0$, and horizontal tangent; (iii) points of ${\mathcal C}$, with $1+dkcos\theta\neq 0$, where the slope of the tangent equals $\displaystyle{\frac{dksin\theta}{1+dkcos\theta}}$.
        \end{itemize}
        \end{theorem}

        {\bf Proof.} From the proof of Theorem \ref{th-no-cusps} it holds that the relationship between the tangents of ${\mathcal C}$ and ${\mathcal G}_{d,\theta}({\mathcal C})$ at corresponding points is $\bar{r}_0'=(I+dkA)\cdot \bar{r}'$. In order to prove (1), one considers the first component of $\bar{r}_0'$, namely $(1+dkcos\theta)x'+dksin\theta y'$, and one imposes that it is $0$. Hence, either $k=0$ and $x'=0$, or $k\neq 0$ and $\displaystyle{\frac{y'}{x'}=-\frac{1+dkcos\theta}{dk sin \theta}}$ (notice that $d\neq 0$ by hypothesis and $sin\theta \neq 0$ because the offset is not classical). Similarly for (2). \qed

\section{Local Shape of the Generalized, Non-classical, Offset} \label{sec-local-shape}

Along this section we consider a real place ${\mathcal P}(h)=(h^p,\beta_q h^q+\xi_r h^r+\cdots)$, non-necessarily regular, a distance $d\neq 0$, and an angle $\theta \neq 0,\pi$ (i.e. we work with a non-classical generalized offset; see \cite{JGS07} for a study of the classical case). Moreover, we write $a=cos \theta$, $b=\sin \theta$, and we represent the coordinates of a place generated by ${\mathcal P}(h)=(x(h),y(h))$ in ${\mathcal G}_{d,\theta}({\mathcal C})$ as $(X(h),Y(h))$. In order to analyze how the generalized offsetting process affects the local shape of ${\mathcal P}(h)$, the idea is to compare the local shape of $(X(h),Y(h))$ with the original local shape. For this purpose, we compute the generalized offset of ${\mathcal P}(h)$ for the previously fixed $d,\theta$. Thus, we have that:
\[
\left(\begin{array}{c}
X(h)\\
Y(h)
\end{array}\right)=
\left(\begin{array}{c}
x(h)\\
y(h)
\end{array}\right)\pm
d\cdot \displaystyle{\frac{1}{\sqrt{x'(h)^2+y'(h)^2}}}\cdot \left(\begin{array}{cc} a & -b \\ b & a \end{array} \right)\cdot \left(\begin{array}{c}
-y'(h) \\ x'(h)
\end{array} \right)
\]
We recall from \cite{JGS07} that, performing computations with formal power series, \[\displaystyle{\frac{1}{\sqrt{x'(h)^2+y'(h)^2}}}=\displaystyle{\frac{1}{h^{p-1}}\cdot \left(\frac{1}{p}-\frac{q^2\beta_q^2}{2p^3}h^{2(q-p)}+\cdots \right)}\]Plugging this expression into the first equality and making computations, one gets that, whenever $\xi_r\neq 0$,
\[X(h)=\mp db+h^p\mp d\displaystyle{\frac{aq\beta_q}{p}h^{q-p}\pm db \frac{q^2\beta_q^2}{2p^2}h^{2(q-p)}\mp d \frac{ar\xi_r}{p}h^{r-p}+\cdots}\]and
\[Y(h)=\pm da\mp d\displaystyle{\frac{bq \beta_q}{p}h^{q-p}\pm d\frac{br\xi_r}{p}h^{r-p}+\cdots}\]Moreover, in the special case when $\xi_r=0$ (i.e. if ${\mathcal P}(h)=(h^p,\beta_q h^q)$) one has that
\[X(h)=\mp db+h^p\mp d\displaystyle{\frac{aq\beta_q}{p}h^{q-p}\pm db \frac{q^2\beta_q^2}{2p^2}h^{2(q-p)}\mp d b\frac{3q^4\beta_q^4}{8p^4}h^{4(q-p)}+\cdots}\]and
\[Y(h)=\pm da\mp d\displaystyle{\frac{bq \beta_q}{p}h^{q-p}\mp da\frac{q^2\beta_q^2}{2p^2}h^{2(q-p)}+\beta_q h^q+\cdots}\]

One may observe that the first order terms of $X(h),Y(h)$ coincide in both cases, $\xi_r\neq 0$ and $\xi_r=0$. Furthermore,
$\mbox{ord}_Y=q-p$. However, $\mbox{ord}_X= \mbox{min}\{p,q-p\}$ and therefore it depends on the sign of $(q-p)-p=q-2p$; moreover, when $q-2p=0$ we also have to distinguish whether the coefficient of  $h^p$ in $X(h)$, namely $1\mp d\displaystyle{\frac{aq\beta_q}{p}}$, is equal to $0$ or not. All these cases ($q-2p>0$, $q-2p=0$, $q-2p<0$) and subcases will be present in our analysis. Furthermore, from Theorem 11 in \cite{JGS07} one may see that the case $q-2p>0$ happens iff the curvature vanishes at the center of the place, while the case $q-2p<0$ occurs iff the curvature tends to infinity as the center of the place is approached.

Also, in the following we separately address results that can be reached by considering only the first order terms of $X(h),Y(h)$ (see Subsection \ref{subsec-first-order}), and results which require to consider also second order terms in $X(h),Y(h)$ (see Subsection \ref{subsec-second-order}). For the second type of results we will need to distinguish the cases $\xi_r\neq 0$ or $\xi_r=0$.

\subsection{Results using a First-order Approximation} \label{subsec-first-order}

We start with the following result; this proposition shows that in some cases, generalized offsetting processes smooth singularities, i.e. they transform singular places into regular ones. This phenomenon happens also for classical offsets (see \cite{JGS07}).

\begin{proposition} \label{regular-places}
Let ${\mathcal P}(h)$ be a place of ${\mathcal C}$ with signature $(p,q)$. If $q-p=1$, then ${\mathcal P}(h)$ generates regular offset places;
as a consequence, if ${\mathcal P}(h)$ is cuspidal (i.e. $p$ is even) and $q-p=1$, then its local shape is not preserved. Conversely, if ${\mathcal P}(h)$ is singular and it is smoothed by the generalized, non-classical, offsetting process (i.e. it generates regular places in the generalized offset), then $q-p=1$.
\end{proposition}

{\bf Proof.} Since $\mbox{ord}_Y=q-p$, if $q-p=1$ we have that the places generated by ${\mathcal P}(h)$ are regular. In particular, in that case these places cannot be cuspidal; so, if ${\mathcal P}(h)$ is cuspidal and $q-p=1$ its local shape is not preserved. Conversely, if ${\mathcal P}(h)$ is singular then $p>1$. Now if it generates regular places then either $\mbox{ord}_X$ or $\mbox{ord}_Y$ is equal to 1. Since $\mbox{ord}_X\geq \mbox{min}\{p,q-p\}$, if $\mbox{ord}_Y=q-p>1$ then $\mbox{min}\{p,q-p\}=1$, which is impossible because both $p,q-p$ are greater than 1. Thus we conclude that $q-p=1$. \qed

Using the results of Section 4 of \cite{JGS07}, one may check that classical offsets also smooth singular places iff $q-p=1$. Now we consider the case $q-2p>0$. In this case, $p<q-p$ and therefore $\mbox{ord}_X=p$. Hence, the following theorem holds.

\begin{theorem} \label{qumenospe}
Let ${\mathcal P}(h)$ be a place of ${\mathcal C}$ with signature $(p,q)$, where $q-2p>0$. Then, the following statements are true:
\begin{itemize}
\item [(i)] If ${\mathcal P}(h)$ is singular, then it generates singular offset places.
\item [(ii)] The local shape of the offset places generated by ${\mathcal P}(h)$ behaves according to the following table:
\begin{center}
\begin{tabular}{c|c|c}
& $p$ even & $p$ is odd \\
\hline
$q$ even  & {\sf thorn} & {\sf flex} \\
\hline
$q$ odd & {\sf beak} & {\sf elbow}
\end{tabular}
\end{center}
\end{itemize}
As a consequence, when $q-2p>0$ the only places whose local shape is preserved are the cuspidal ones.
\end{theorem}

{\bf Proof.} Since $q-2p>0$, then $p<q-p$ and $\mbox{ord}_X=p$; moreover, since $\mbox{ord}_Y=q-p$ then $\mbox{ord}_X<\mbox{ord}_Y$. Hence, the signature of an offset place generated by ${\mathcal P}(h)$ is $(p_0,q_0)=(p,q-p)$. Now if ${\mathcal P}(h)$ is singular then $p>1$; therefore $p_0>1$ and the offset place is singular. Moreover, the above table is also derived from the fact that $(p_0,q_0)=(p,q-p)$
. From this table one may deduce that the local shape is preserved iff $p$ is even. \qed

\begin{remark} \label{no-contradiction}
Notice that when $q-p=1$, $q-2p=1-p$ and since $p\geq 1$, it holds that $q-2p\leq 0$; hence, the case $q-2p>0$ cannot occur and we find no contradiction between the first statement of Theorem \ref{qumenospe} and Proposition \ref{regular-places}.
\end{remark}

So, we see that the case $q-2p>0$ is completely described just by using first order terms. When $q-2p\leq 0$, the orders of $X(h)$ and $Y(h)$ are in general both equal to $q-p$; so, denoting as $(p_0,q_0)$ the signature of a place generated by ${\mathcal P}(h)$, we have that while $p_0=q-p$, in order to compute $q_0$ we need to consider higher order terms. One may see that this situation is quite different from the classical one, where first order terms are enough to provide a good description of the cases $q-2p=0$ and $q-2p<0$ (see \cite{JGS07}). Nevertheless, using just the relationship $p_0=q-p$, the following result concerning the case $q-2p<0$ can be derived.

\begin{proposition} \label{p-cero}
Let ${\mathcal P}(h)$ be a place of ${\mathcal C}$ with signature $(p,q)$, where $q-2p< 0$. If $q$ is odd, then the local shape of ${\mathcal P}(h)$ is not preserved.
\end{proposition}

{\bf Proof.} Since $p_0=q-p$ then if $p$ is even and $q$ is odd, $q-p$ is odd and the local shape is not preserved. On the other hand, if $p,q$ are both odd then $q-p$ is even and the local shape is not preserved, either. \qed

Theorem \ref{qumenospe} and Proposition \ref{p-cero} provide the following corollary on the non-preservation of the flex points of ${\mathcal C}$.

\begin{corollary} \label{no-flex-points}
The generalized, non-classical, offset never preserves flex points.
\end{corollary}

{\bf Proof.} Let ${\mathcal P}(h)$ be a real place with signature $(p,q)$, whose center is a flex point. Then, from Definition \ref{def1} $p,q$ are both odd. Hence $q$ cannot be equal to $2p$, i.e. either $q-2p>0$ or $q-2p<0$ hold. In the first case, the result follows from Theorem \ref{qumenospe}; in the second case, the result follows from Proposition \ref{p-cero}. \qed

In order to give a more complete description of the cases $q-2p=0$ and $q-2p<0$, we need to take into account higher order terms in $X(h)$. This is considered in the next subsection.

\begin{example}
Consider the curve $x^3-y^2=0$, and the place ${\mathcal P}(h)=(h^2,h^3)$ centered at the origin. Here we have $p=2,q=3$, and therefore $q-2p<0$. Since $q$ is odd, from Proposition \ref{p-cero} we deduce that the local shape of ${\mathcal P}(h)$ is not preserved by any generalized offset. In fact, since this place is cuspidal and $q-p=1$, from Proposition \ref{regular-places} it follows that ${\mathcal P}(h)$ generates regular offset places. In Figure 5 one may see (in thick line) the generalized offset to the curve for $d=1$ and $\theta=\pi/4$; here one may check that the generalized offset contains no cusp.

\begin{figure}[ht]
\begin{center}
\centerline{\psfig{figure=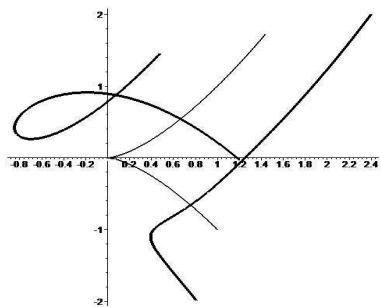,width=5cm,height=4cm}}
\caption{Generalized Offset to $x^3-y^2=0$, $\theta=\pi/4$, $d=1$}
\end{center}
\end{figure}
\end{example}





\subsection{Results using a Second Order Approximation}\label{subsec-second-order}

In this section we provide a more complete description of the phenomenon when $q-2p\leq 0$. For this purpose, we consider a second order approximation of $X(h), Y(h)$. Furthermore, in the following we analyze in detail the case $q-2p=0$. The analysis of the case $q-2p<0$ is similar; so, for this other case we give the results without proofs, leaving them to the reader.

\subsubsection{The case $q-2p=0$}\label{subsubsec-q-2p}

We start assuming that $\xi_r\neq 0$; the case $\xi_r=0$ will be addressed at the end of the subsection. Now in this case we have that \[X(h)= \left(1\mp d \displaystyle{\frac{aq\beta_q}{p}}\right) h^p+\cdots\]and therefore we have to distinguish whether $1\mp d \displaystyle{\frac{aq\beta_q}{p}} \neq 0$, or not; in the first case $\mbox{ord}_X=p$, while in the second case $\mbox{ord}_X>p$. Furthermore, the following lemma, concerning the curvature at the center of the considered place, will be useful. Here we recall that a place ${\mathcal P}(h)$ can be taken as a parametrized curve for $h\in I$, where $I$ is an interval containing $0$ where the components of the place converge. So, plugging the coordinates of the place into the curvature formula one obtains the function curvature. If ${\mathcal P}(h)$ is regular, then the resulting function is analytic in $I$. The following lemma takes into consideration not only this situation, but also the alternative one which arises when ${\mathcal P}(h)$ is singular.

\begin{lemma} \label{curvature}
Let ${\mathcal P}(h)=(h^p,\beta_qh^q+\xi_r h^r+\cdots)$ be a real place of ${\mathcal C}$ with $q-2p=0$, and let $P$ be the center of ${\mathcal P}(h)$. Then, the function curvature $k_h$ of ${\mathcal P}(h)$ satisfies that:
\begin{itemize}
\item [(1)] If $p=1$ (i.e. the place is regular), then $k_h=\displaystyle{2\beta_qh^{q-2}+\cdots}$.
\item [(2)] If $p>1$ (i.e. the place is singular), then
\[k_h=
\left\{\begin{array}{lrc}
\displaystyle{2\beta_q +\frac{r(r-p)\xi_r}{p^2}h^{r-2p}+\cdots} & \mbox{if} & h>0 \\
\displaystyle{-2\beta_q -\frac{r(r-p)\xi_r}{p^2}h^{r-2p}+\cdots} & \mbox{if} & h<0
\end{array}\right.
\]As a consequence, $k_h$ and the derivative $k_h^{(r-2p)}$ are not continuous at $h=0$; however, $|k_h|$ and $|k_h^{(r-2p)}|$ have a removable discontinuity at $h=0$ and therefore they can be extended to functions $\hat{k}_h$, $\hat{m}_h$, respectively, continuous at $h=0$. In particular, these functions satisfy $\hat{k}_h(0)=2|\beta_q|$, $\hat{m}_h(0)=\displaystyle{\frac{r(r-p)(r-2p)!|\xi_r|}{p^2}}$.
\end{itemize}
\end{lemma}

{\bf Proof.} The above expression for $k_h$ can be obtained by plugging the coordinates of ${\mathcal P}(h)$ into the curvature formula and doing computations with formal power series (see \cite{JGS07}), taking into account that $(x'^2(h)+y'(h)^2)^{3/2}=|h^{3p-3}|\cdot (p^2+{\mathcal O}(h^{2p}))^{3/2}$. For the second statement one studies limits at the right and at the left of $h=0$. \qed


In the following we will use the notation $\tilde{k}=2\beta_q$, and $\tilde{m}=\displaystyle{\frac{r(r-p)(r-2p)!\xi_r}{p^2}}$; since in this subsection we are working with a place ${\mathcal P}(h)$ satisfying that $q-2p=0$, from Lemma \ref{curvature} these quantities correspond to the right limits of the curvature and of the $(r-2p)$-derivative of the curvature, respectively, of ${\mathcal P}(h)$ as one approaches its center. Using this notation,the expression $1\mp d \displaystyle{\frac{aq\beta_q}{p}}$ is equivalent to $\pm \tilde{k}=\displaystyle{\frac{1}{da}}$. Hence, the following theorem holds.

\begin{theorem}\label{first-case-q-2p=0}
Let ${\mathcal P}(h)=(h^p,\beta_qh^q+\xi_r h^r+\cdots)$ be a real place of ${\mathcal C}$ satisfying that $q-2p=0$. If $\tilde{k}=\mp 1/da$, then the following behavior is obtained.
\begin{itemize}
\item [(1)] $r>3p$: preserved.
\item [(2)] $r<3p$: preserved if and only if $r,p$ are both even
or both odd.
\item [(3)] $r=3p$: if $\displaystyle{\frac{b}{2}\cdot \tilde{k}^2-\frac{ap}{r-p}\cdot
\frac{\tilde{m}}{(r-2p)!}\neq 0}$, preserved.
\end{itemize}
\end{theorem}

{\bf Proof.} Since $\tilde{k}=\mp 1/da$, the coefficient of $h^{p}$ in the $x$-coordinate $X(h)$ of one of the offset places generated by ${\mathcal P}(h)$, vanishes. Hence, for that place it holds that $\mbox{ord}_X=\mbox{min}\{2(q-p)=2p,r-p\}$. Thus, in order to compute $\mbox{ord}_X$ we have to discuss whether $2p$ is greater than $r-p$ or not (i.e. whether $r$ is greater than $3p$); moreover, in case that $r=3p$, we also have to analyze whether the coefficient of $h^{2p}$ vanishes or not. Now let $(p_0,q_0)$ be the signature of the offset place. Since $q-2p=0$ it follows that $\mbox{ord}_Y=q-p=p$. Then if $r>3p$ we have that $\mbox{ord}_Y=p<\mbox{ord}_X=2p$, and therefore $(p_0,q_0)=(p,2p)=(p,q)$; hence, the local shape is preserved. If $r<3p$ then $\mbox{ord}_Y=q-p=p<\mbox{ord}_X=r-p$ (notice that since $p<q$ and $q<r$, then $q-p=p<r-p$); hence, $(p_0,q_0)=(p,r-p)$ and the local shape is preserved iff $r,p$ are both even or both odd. Finally, if $r=3p$ then \[X(h)=\pm d \cdot \displaystyle{\frac{1}{p}\cdot \left(\frac{bq^2\beta_q^2}{2p}-ar\xi_r\right)} h^{2p} + \cdots\]
From Lemma \ref{curvature}, one may check that the coefficient of $h^{2p}$ in $X(h)$ vanishes iff $\displaystyle{\frac{b}{2}\cdot \tilde{k}^2-\frac{ap}{r-p}\cdot
\frac{\tilde{m}}{(r-2p)!}=0}$. Thus, if this does not happen then $\mbox{ord}_X=2p=q$ and therefore $(p_0,q_0)=(p,q)$; so, the local shape is preserved. \qed

Similarly the following theorem holds.

\begin{theorem}\label{second-case-q-2p=0}
Let ${\mathcal P}(h)=(h^p,\beta_qh^q+\xi_r h^r+\cdots)$ be a real place of ${\mathcal C}$ satisfying that $q-2p=0$. If $\tilde{k}\neq \pm 1/da$, then:
\begin{itemize}
\item [(1)] $r>3p$: if $a\mp d\tilde{k}(a^2-b^2)\neq 0$, preserved.
\item [(2)] $r=3p$: if $\displaystyle{\frac{\tilde{k}^2}{2}(a\mp
d\tilde{k}(a^2-b^2))-\frac{pb}{r-p}\cdot \frac{\tilde{m}}{(r-2p)!}\neq
0}$, preserved.
\item [(3)] $r<3p$: preserved if and only if $r,p$ are both even
or both odd.
\end{itemize}
\end{theorem}

{\bf Proof.} We may observe that in this case $\mbox{ord}_X=\mbox{ord}_Y=q-p=p$. More precisely, it holds that
\[
\begin{array}{l}
X(h)=\displaystyle{\mp db+\left(1\mp d\cdot \frac{a\beta_q}{p}\right)h^p\mp db \cdot \frac{q^2\beta_q^2}{2p^2}h^{2(q-p)}\mp d\cdot \frac{ar\xi_r}{p}h^{r-p}+\cdots}\\
Y(h)=\displaystyle{\pm da \mp d\cdot \frac{bq\beta_q}{p}h^p\pm d \cdot \frac{br\xi_r}{p}h^{r-p}\mp da \frac{q^2\beta_q^2}{2p^2}h^{2(q-p)}+\cdots}
\end{array}
\]These expressions can be written as
\[
\begin{array}{l}
X(h)=u_0+u_1h^p+bBh^{2(q-p)}+aCh^{r-p}+\cdots\\
Y(h)=v_0+v_1h^p+aBh^{2(q-p)}+bCh^{r-p}+\cdots
\end{array}
\]where $u_0=\mp db$, $v_0=\pm da$, $B=\mp db \cdot \displaystyle{\frac{q^2\beta_q^2}{2p^2}}$, etc. Now like in the previous theorem, we have to discuss the value of $\mbox{min}\{2(q-p)=2p,r-p\}$, which is equivalent to discussing the value of $\mbox{min}\{r,3p\}$. So, let us consider first that $\mbox{min}\{2(q-p)=2p,r-p\}=2(q-p)$, i.e. that $r>3p$. Then, we can compute the local shape of the place by directly applying Definition
\ref{pairassociated}. For this purpose, we write ${\mathcal Q}(h)=(X(h),Y(h))$ and we represent by $(p_0,q_0)$ its signature. Clearly $p_0=p$; moreover, ${\mathcal Q}^{(p)}(h)$ is parallel to $(u_1,v_1)$. So, in order to determine $q_0$ we have to find the least natural number, greater than $p$, so that ${\mathcal Q}^{(p)}(h)$ and ${\mathcal Q}^{(k)}(h)$ are linearly independent. For $k\in (p,2(q-p))$ it holds that ${\mathcal Q}^{(p)}(k)=\vec{0}$. Hence, the smallest possible value for $q_0$ is $2(q-p)=2p=q$. Moreover, since $B\neq 0$, one may check that ${\mathcal Q}^{(2(q-p))}(h)$ is parallel to $(a,b)$. Then, if $(u_1,v_1)$ and $(a,b)$ are not parallel, i.e. if $u_1a-v_1b\neq 0$, it holds that  $q_0=2(q-p)=q$, and hence the local shape is preserved. Making computations, one can check that the inequality $u_1a-v_1b\neq 0$ is equivalent to $a\mp d\tilde{k}(a^2-b^2) \neq 0$, i.e. $\tilde{k}\neq \pm a/d(a^2-b^2)$. So, the first statement follows. For the other cases $r=3p$, $r<3p$ one also applies Definition
\ref{pairassociated} and similar reasonings. \qed

\begin{example}
Consider the curve of equation $x^9-y^2+2yx^2-x^4=0$, which contains the origin. A place of this curve centered at the origin is
${\mathcal P}(h)=(h^2,h^4+h^9)$, which satisfies $p=2,q=4$ and therefore $q-2p=0$; moreover, $r=9>3p=6$. Also, one may see that the absolute value of the curvature of ${\mathcal P}(h)$ at $h=0$ is $|k|=2$. Hence, from Theorem \ref{first-case-q-2p=0} and Theorem \ref{second-case-q-2p=0} it follows that in the following cases, the local shape is preserved: (i) when $da\neq \pm 1/2$; (ii) when $da=\pm 1/2$, and $a\mp 2d(a^2-b^2)\neq 0$. For example, in Figure 6 one has (in thick line) the generalized offset for $d=1$ and $\theta=\pi/4$. Here  $a=b=\sqrt{2}/2$, and therefore $a\mp 2d(a^2-b^2)=a\neq 0$; in particular, one may check that the local shape has been preserved (i.e. the cusp in the original curve has generated two offset cusps of the same type).

\begin{figure}[ht]
\begin{center}
\centerline{\psfig{figure=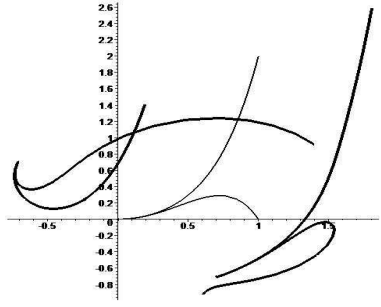,width=5cm,height=4cm}}
\caption{Generalized Offset to $x^9-y^2+2yx^2-x^4=0$, $\theta=\pi/4$, $d=1$}
\end{center}
\end{figure}
\end{example}


Finally, the above results hold whenever $\xi_r\neq 0$. So, let us briefly address the case when $\xi_r=0$. In this case, we have that ${\mathcal P}(h)=(h^p,\beta_q h^q)$, with $q=2p$. Hence, changing the parameter we can write the place as ${\mathcal P}(\bar{h})=(x(\bar{h}),y(\bar{h}))$, and we see that it corresponds to a regular place locally describing a parabola. Hence, by Theorem \ref{th-reg-flex-points} in Section \ref{sec-behav-reg}, it gives rise to regular offset places. Furthermore, by applying Theorem \ref{th-reg-flex-points} (or equivalently by doing computations with places), one may see that the only cases when the local shape may not be preserved fulfill $1\mp 4da\beta_q+4d^2\beta_q^2=0$; in this situation, flex points may arise.

\subsubsection{The case $q-2p<0$}\label{subsubsec-q-menos-2p}

In this section we provide the results without proofs; these are tedious and similar to those in the preceding subsection, and are left to the reader. Moreover, for simplicity here we use the notation  $\tilde{u}=q\beta_q/p$, and $\tilde{v}=\displaystyle{\frac{r(r-p)(r-2p)!\xi_r}{p^2}}$, analogous to the notation introduced in the preceding section. However, unlike in the case $q-2p=0$, here these quantities do not have any specific geometrical meaning. Finally, notice also that when $q-2p<0$ the center of the place is singular; indeed, for regular places $p=1$, and since $q>p$ one gets that $q-2p\geq 0$. So, the case $q-2p<0$ is only concerned with singularities.

We consider first the special case when $\xi_r=0$. In this case, the following theorem holds.

\begin{theorem} \label{second-res}
Let ${\mathcal P}(h)=(h^p,\beta_qh^q)$ be a real place of ${\mathcal C}$ satisfying that $q-2p<0$. If $p\neq 2(q-p)$ or $p=2(q-p)$ but $1\pm db\tilde{u}^2/2\neq 0$, $\pm b+\tilde{u}^2d/2=0$, then the local shape of ${\mathcal P}(h)$ is preserved.
\end{theorem}

In the more general case $\xi_r\neq 0$, the following result holds.

\begin{theorem} \label{first-result-q-2p-neg}
Let ${\mathcal P}(h)=(h^p,\beta_qh^q+\xi_r h^r+\cdots)$ be a real place of ${\mathcal C}$ satisfying that $q-2p<0$ and $\xi_r\neq 0$. Then, the local behavior of ${\mathcal P}(h)$ verifies the following:
\begin{itemize}
\item [(1)] If $2(q-p)<r-p$, then:
\begin{itemize}
\item [a.] If $p<2(q-p)$ the local shape is preserved iff $p,q$ are both even.
\item [b.] If $p\geq 2(q-p)$ the local shape is preserved iff $q$ is even.
 \end{itemize}
 \item [(2)] If $2(q-p)=r-p$, then:
 \begin{itemize}
 \item [a.] If $p<2(q-p)$ the local shape is preserved iff $p,q$ are both even.
 \item [b.] If $p>2(q-p)$ then:
 \begin{itemize}
 \item [b.1] If $\displaystyle{\frac{bp}{r-p}\frac{\tilde{v}}{(r-2p)!}-\frac{a}{2}\tilde{u}^2\neq 0}$ and $\displaystyle{\frac{b^2-a^2}{2}\tilde{u}^2-\frac{abp}{r-p}\frac{\tilde{v}}{(r-2p)!}\neq 0}$, then the local shape is preserved iff $q$ is even.
     \item [b.2] If $\displaystyle{\frac{bp}{r-p}\frac{\tilde{v}}{(r-2p)!}-\frac{a}{2}\tilde{u}^2=0}$, the local shape is preserved iff $q$ is even.
         \end{itemize}
         \item [(c)] If $p=2(q-p)$ then:
         \begin{itemize}
         \item [c.1] If $\displaystyle{1\pm db\frac{1}{2}\tilde{u}^2\mp da \frac{p}{r-p}\frac{\tilde{v}}{(r-2p)!}\neq 0}$ and $\displaystyle{\pm b\tilde{u}+d\frac{1}{2}(a^2-b^2)\tilde{u}^2-2da\frac{bp}{r-p}\frac{\tilde{v}}{(r-2p)!}\neq 0}$, the local shape is preserved iff $q$ is even.
             \item[c.2] If $\displaystyle{1\pm db\frac{1}{2}\tilde{u}^2\mp da \frac{p}{r-p}\frac{\tilde{v}}{(r-2p)!}= 0}$ the local shape is preserved iff $q$ is even.
                 \end{itemize}
                 \end{itemize}
                 \item [(3)] If $2(q-p)>r-p$, then:
                 \begin{itemize}
                 \item [a.] If $p<r-p$, then the local shape is preserved iff $p,q$ are both even.
                 \item [b.] If $\theta\neq \pi/2$, then the local shape is preserved iff $q$ is even and $r,p$ are both even or both odd.
                     \item [c.] If $p=r-p$, then if either $\displaystyle{1\mp da \frac{p}{r-p}\frac{\tilde{v}}{(r-2p)!}=0}$, or $\displaystyle{1\mp da \frac{p}{r-p}\frac{\tilde{v}}{(r-2p)!}\neq 0}$ and $\displaystyle{(da\tilde{u}\pm 1)\cdot \frac{p}{r-p}\frac{\tilde{v}}{(r-2p)!}\neq 0}$, the local shape is preserved iff $q$ is even and $r,p$ are both even or both odd.
                         \end{itemize}
                         \end{itemize}
                         \end{theorem}

\section{Conclusions and Comparison between Classical and Non-Classical Generalized Offsets} \label{sec-comparison}

In the preceding sections we have analyzed local aspects on the shape of generalized offsets, both using tools coming from Differential Geometry and using the notion of local shape. In this section, we summarize the main results we have obtained in our analysis, and we compare them with the local properties on the shape of classical offsets that are derived in \cite{JGS07} and \cite{Faroukki}.

Now the following table summarizes the most relevant properties concerning local aspects of the classical offset shape; we refer the reader to \cite{JGS07}, \cite{Faroukki} for further reading on them.

\begin{center}
\begin{tabular}{c|c}
& Classical Offsets  \\
\hline
Regular Points & $\begin{array}{l} \mbox{Generate singular places when }k=-1/d \mbox{; cusps may arise} \\ \mbox{Flex points are preserved } \\ \mbox{Turning points are preserved} \\ \mbox{Tangents preserved} \end{array} $ \\
\hline
Singular Points & $ \begin{array}{l} \mbox{Smoothed iff } q-p=1 \\ \mbox{Singular flex points preserved when } q-2p>0 \\ q-2p>0: \mbox{ preserved}  \\ q-2p=0:  \mbox{ if } |k|\neq 1/d \mbox{, preserved} \\ q-2p<0: \mbox{ preserved iff } q \mbox{ is even.}
\end{array}$ \\
\end{tabular}
\end{center}

The following table shows analogous properties for the generalized, non-classical offset; these properties are derived from the results in this paper.

\begin{center}
\begin{tabular}{c|c}
& Non-Classical Generalized Offsets  \\
\hline
Regular Points & \begin{tabular}{l} Never generate singular places; cusps do not arise. \\ Flex points are never preserved. \\ Turning points are not preserved in general. \\ Tangents not preserved. \end{tabular} \\
\hline
Singular Points & \begin{tabular}{l} Smoothed iff $q-p=1$. \\ Singular flex points never preserved. \\ $q-2p>0:$ preserved iff $p$ even \\ $q-2p=0:$ distinguish $|k|\neq 1/(d cos \theta)$, or not; many subcases. \\ $q-2p<0:$ many subcases.
\end{tabular} \\
\end{tabular}
\end{center}

Hence, we observe a great number of differences between the local behavior in the classical and the non-classical case, both at regular and singular points (where the situation is far more intricate in the non-classical case).


\begin{thebibliography}{56}

\bibitem{JGglobal} Alcazar J.G. (2008) {\it Good Global Behavior of Offsets to Plane Algebraic Curves}, Journal of Symbolic Computation vol. 43, pp. 659-680.

\bibitem{JG06} Alcazar J.G. (2008) {\it Local Shape
of Offsets to Implicit Algebraic Curves}, submitted.





\bibitem{JGS05} Alcazar J.G., Sendra J.R. (2006) {\it Local Shape
of Offsets to Rational Algebraic Curves}, Tech. Report SFB 2006-22
(RICAM, Austria)

\bibitem{JGS07} Alcazar J.G., Sendra J.R. (2007) Alcazar J.G., Sendra R. (2007) {\it Local Shape
of Offsets to Algebraic Curves}, Journal of Symbolic Computation
vol. 42, pp. 338-351.

\bibitem{ASS96} Arrondo E., Sendra J., Sendra J.R. (1997).
{\it  Parametric Generalized Offsets to Hypersurfaces}.   Journal
of Symbolic Computation vol. 23, pp. 267--285.

\bibitem{ASS97} Arrondo E., Sendra J., Sendra J.R. (1999).
{\it   Genus Formula for Generalized Offset Curves}, Journal of
Pure and Applied Algebra vol. 136, no. 3, pp. 199--209.




\bibitem{Farin} Farin G., Hoscheck J., Kim M-S. (2002). {\it
Handbook of Computer Aided Geometric Design}, North-Holland.

\bibitem{Faroukki} Farouki R.T., Neff C.A. (1990). {\it Analytic
Properties of Plane Offset Curves}, Computer Aided Geometric
Design vol. 7, pp. 83--99.

\bibitem{Far2} Farouki R.T., Neff C.A. (1990). {\it Algebraic
Properties of Plane Offset Curves}, Computer Aided Geometric
Design vol. 7, pp. 101--127.









\bibitem{HL97} Hoschek J., Lasser D. (1993), {\it Fundamentals
of    Computer Aided Geometric Design}.  A.K. Peters Wellesley
MA., Ltd.




\bibitem{PP98b} Pottmann H., Peternell M. (1998), {\it
A Laguerre Geometric Approach to Rational Offsets}.  Computer
Aided Geometric Design vol. 15,  223-249.





\bibitem{Juani-thesis} Sendra J. (1999) {\it Algoritmos efectivos para la
manipulacion de offsets de hipersuperficies}, PhD Thesis,
Universidad Politecnica de Madrid.


\bibitem{SS99}   Sendra J., Sendra J.R.  (2000).
{\it Algebraic Analysis of Offsets to Hypersurfaces}.
Mathematische Zeitschrift vol. 234, pp. 697--719.

\bibitem{SS00}   Sendra J., Sendra J.R.  (2000).
{\it Rationality Analysis and Direct Parametrization of
Generalized Offsets to Quadrics}. Applicable Algebra in
Engineering, Communication and Computing  vol. 11, no. 2, pp.
111--139.




\bibitem{walker} Walker R.\,J. (1950). {\it  Algebraic Curves.}
Princeton University Press, Princeton.
\end{thebibliography}
\end{document}